# Mixed spin (1/2-1) hexagonal Ising nanowire: some dynamic behaviors

Ersin Kantar[1, *] and Yusuf Kocakaplan[2]
[1]*Department of Physics, Erciyes University, 38039 Kayseri, Turkey*
[2]*Graduate School of Natural and Applied Sciences, Erciyes University, 38039 Kayseri, Turkey*

**Abstract**

The dynamic behaviors of a mixed spin (1/2-1) hexagonal Ising nanowire (HIN) with core-shell structure in the presence of a time dependent magnetic field are investigated by using the effective-field theory with correlations based on the Glauber-type stochastic dynamics (DEFT). According to the values of interaction parameters, temperature dependence of the dynamic magnetizations, the hysteresis loop areas and the dynamic correlations are investigated to characterize the nature (first- or second-order) of the dynamic phase transitions (DPTs). Dynamic phase diagrams, including compensation points, are also obtained. Moreover, from the thermal variations of the dynamic total magnetization, the five compensation types can be found under certain conditions, namely the Q-, R-, S-, P-, and N-types.

***Keywords:*** Hexagonal Ising nanowire, Effective-field theory, Glauber-type stochastic dynamics, Dynamic phase diagram, Compensation behavior.

## 1. Introduction

Recently, magnetic nanomaterials (nanoparticles, nanofilms, nanorods, nanobelts, nanowires and nanotubes etc.) have attracted a great interest both theoretically and experimentally. The reason is that these materials can be used technological area, such as biomedical applications [1, 2], sensors [3], nonlinear optics [4], permanent magnets [5], environmental remediation [6], and information storage devices [7–9]. In particular, magnetic nanowire systems have attracted considerable attention not only because of their academic interest, but also the technological applications. In the experimental area, the magnetic nanowires have been synthesized and their magnetic properties have been investigated, such as Fe–Co [10], Co–Pt [11], Ni [12], Ga1-xCuxN [13], Fe [14], Fe3O4 [15], Co [16], Fe–Pt [17], Ni-Fe [18], Co-Cu [19] etc. In theoretical area, the magnetic nanowires have been investigated within the various theoretical methods, such as effective-field theory (EFT) with correlations [20-24], Monte Carlo Simulations (MCs) [25].

On the other hand, the mixed spin Ising systems have attracted a great deal of attention and intensively investigated within the concept of statistical physics. These systems have observed many new phenomena that cannot be exhibit in single-spin Ising systems. The most extensively mixed system is mixed spin (1/2-1) Ising system. This model has been studied by the mean-field approximation (MFA) [26–29], MCs [30, 31] and EFT with correlation [32-35]. Moreover, some further works about magnetic nanomaterials of mixed spin (1/2-1) Ising system were given [36-40].

---

[*]Corresponding author.
Tel: + 90 352 2076666 # 33136
*E-mail address:* ersinkantar@erciyes.edu.tr (Ersin Kantar)

Finally, we should also mentioned that the dynamic phase transition (DPT) temperature has attracted much attention in recent years, both theoretically [41–49] and experimentally [50–54]. However, as far as we know, the DPT temperatures of the magnetic nanostructured materials have only been investigated a few works by usin EFT [55-58] and MCs [59-60]. Therefore, the aim of this paper is to investigate temperature dependence of the dynamic magnetizations and the dynamic phase diagrams, including compensation points, of the HIN in an oscillating magnetic field within the DEFT.

The paper is organized as follows. In Section II, we define the model and give briefly the formulation of a mixed spin (1/2-1) HIN by using the DEFT. In Section III, we present the numerical results and discussions. Finally, Section IV contains the summary and conclusions.

## 2. Model and Formulation

The Hamiltonian of the hexagonal Ising nanowire (HIN) includes nearest neighbor interactions and the crystal field is given as follows:

$$H = -J_S \sum_{\langle ij \rangle} S_i S_j - J_C \sum_{\langle mn \rangle} \sigma_m \sigma_n - J_1 \sum_{\langle im \rangle} S_i \sigma_m - D \sum_i S_i^2 - h(t) \left( \sum_i S_i + \sum_m \sigma_m \right) \qquad (1)$$

where $\sigma = \pm 1/2$ and $S = \pm 1, 0$. The $J_S$, $J_C$ and $J_1$ are the exchange interaction parameters between the two nearest-neighbor magnetic particles at the shell surface, core and between the shell surface and core, respectively (see Fig. 1). D is a Hamiltonian parameter and stands for the single-ion anisotropy (i.e. crystal field). The surface exchange interaction $J_S = J_C(1+\Delta_S)$ and interfacial coupling $r = J_1 / J_C$ are often defined to clarify the effects of the surface and interfacial exchange interactions on the physical properties in the nanosystem, respectively. h(t) is a time-dependent external oscillating magnetic field and is given as $h(t) = h_0 \sin(wt)$. In here, $h_0$ and $\omega = 2\pi\nu$ are the amplitude and the angular frequency of the oscillating field, respectively. The system is in contact with an isothermal heat bath at absolute temperature $T_A$.

For the mixed spin (1/2-1) HIN system, within the framework of the EFT with correlations, one can easily find the magnetizations $m_S$, $m_C$ and quadrupole moment $q_S$ as coupled equations as follows:

$$\begin{Bmatrix} m_S \\ q_S \end{Bmatrix} = \left[1 + m_S \sinh(J_S \nabla) + m_S^2 (\cosh(J_S \nabla) - 1)\right]^4 \left[\cosh(J_1 \nabla/2) + 2m_C \sinh(J_1 \nabla/2)\right] \begin{Bmatrix} F_m(x)|_{x=0} \\ F_q(x)|_{x=0} \end{Bmatrix}, \qquad (2a)$$

$$m_C = \left[\cosh(J_C \nabla/2) + 2m_C \sinh(J_C \nabla/2)\right]^2 \left[1 + m_S \sinh(J_1 \nabla) + m_S^2 (\cosh(J_1 \nabla) - 1)\right]^6 G_m(x)|_{x=0}, \qquad (2b)$$

where $\nabla = \partial/\partial x$ is the differential operator. The functions F(x) and G(x) are defined as

$$F_m(x) = \frac{2\sinh[\beta(x+h)]}{\exp(-\beta D)+2\cosh[\beta(x+h)]} \tag{3a}$$

$$F_q(x) = \frac{2\cosh[\beta(x+h)]}{\exp(-\beta D)+2\cosh[\beta(x+h)]} \tag{3b}$$

$$G_m(x) = \frac{1}{2}\tanh\left[\frac{1}{2}\beta(x+h)\right]. \tag{3c}$$

where $\beta = 1/k_B T_A$, $T_A$ is the absolute temperature and $k_B$ is the Boltzman factor.

In this point, we can obtain the set of the dynamical effective-field equations by means of the Glauber-type stochastic dynamics. We employ the Glauber transition rates, which the system evolves according to the Glauber-type stochastic process at a rate of $1/\tau$ transitions per unit time. Hence, the frequency of spin flipping, $f$, is $1/\tau$. After some manipulations the set of dynamic equations of motion for the magnetizations are obtained as:

$$\tau \frac{dm_S}{dt} = -m_S + f(m_S, m_C) \tag{4a}$$

$$\tau \frac{dm_C}{dt} = -m_C + g(m_C, m_S), \tag{4b}$$

In here, f and g functions came from the expanding right-hand side of Eqs. 2(a) and 2(b), respectively. These functions consist long coefficients that and can be easily calculated by employing differential operator technique, namely $exp(a\nabla)f(x) = f(x+a)$. But, these coefficients will not be expressed here because of complicate and long expressions. The dynamic order parameters or dynamic magnetizations as the time-averaged magnetization over a period of the oscillating magnetic field are given as

$$M_\alpha = \frac{w}{2\pi}\oint m_\alpha(t)\,dt, \tag{5}$$

where α = S (Shell), C (Core), T (Total) which correspond to the dynamic magnetizations for the shell, core, and the dynamic total magnetization, respectively. On the other hand, the hysteresis loop area is defined by Acharyya [43] as

$$A_\alpha = -\oint m_\alpha(t)\,dh = -h_0 w \oint m_\alpha(t)\cos(wt)\,dt, \tag{6}$$

which corresponds to the energy loss due to the hysteresis. The dynamic correlations are calculated as

$$C_\alpha = \frac{w}{2\pi}\oint m_\alpha(t)\,h(t)\,dt = \frac{wh_0}{2\pi}\oint m_\alpha(t)\sin(wt)\,dt. \tag{7}$$

We should also mention that in the numerical calculations, the hysteresis loop areas $A_\alpha$ and the dynamic correlations $C_\alpha$ are also measure in units $J_C$. The dynamic total magnetization $M_T$ vanishes at the compensation temperature $T_{comp}$. Then, the compensation point can be determined by looking for the crossing point between the absolute values of the surface and the core magnetizations. Therefore, at the compensation point, we must have

$$\left| M_S \left( T_{comp} \right) \right| = \left| M_C \left( T_{comp} \right) \right|, \tag{8}$$

and

$$\text{sgn}\left[ M_S \left( T_{comp} \right) \right] = -\text{sgn}\left[ M_C \left( T_{comp} \right) \right]. \tag{9}$$

We also require that $T_{comp} < T_C$, where $T_C$ is the critical point temperature. In the next section we will give the numerical results of these equations.

### 3. Numerical Results and Discussions

In this section, we investigate behavior of time variations of average order parameters to find phases in this system. Then, we calculated phase diagrams in the different planes, namely the (T, h), (D, T), (r, T) and ($\Delta_S$, T) planes. Finally, we obtain the dynamic compensation points and determine different dynamic compensation behaviors. We has fixed $J_C = 1.0$ throughout of the paper.

*3.1. The phases in the system: Time variations of average order parameters*

At first, the time variations of the average shell and core magnetizations are investigated to obtain the phases in the system. In order to determine the behaviors of time variations of the average magnetizations, the stationary solutions of the dynamic effective-field coupled equations, namely Eqs. (4a)-(4b), have been studied for various values of the system parameters. The stationary solutions of these equations will be a periodic function of $\xi$ with period $2\pi$; that is, $m_S(\xi+2\pi) = m_S(\xi)$ and $m_C(\xi+2\pi) = m_C(\xi)$. Moreover, they can be one of the three types according to whether they have or do not have the properties

$$m_S(\xi+\pi) = -m_S(\xi), \tag{10a}$$

and

$$m_C(\xi+\pi) = -m_C(\xi). \tag{10b}$$

where $\xi = \omega t$. By utilizing the Adams-Moulton predictor-corrector method, we can solve Eqs. (10a) and (10b) for a given set of parameters and initial values. The first type solution of Eqs. (10a) and (10b) is a symmetric solution and it corresponds to a paramagnetic (p) phase. In the symmetric solution, average shell and core magnetizations delayed with respect to the external magnetic field. The second type solution of Eqs. (10a) and (10b) is called a non-symmetric solution that corresponds to a ferrimagnetic (i) solution. In this case, average shell and core magnetizations do not follow the external magnetic field any more, but instead of oscillating around zero value. The results of these solutions are presented in Fig. 2. Fig. 2(a)-(c) display p, i and nonmagnetic (nm) fundamental phases for different physical parameters and initial values, respectivelly. In Fig. 2(a), the initial values of average shell magnetization $m_S$ = 1.0 and -1.0, and average core magnetization $m_C$ = 0.5 and -0.5 and oscillate around zero value, namely $m_C(\xi) = m_S(\xi) = 0$. Hence, the system shows symetric solution, namely p phase. In Fig. 2(b), average shell and core magnetizations have different initial values. The shell magnetization oscillates around 1.0 value, core magnetization oscillates around 0.5 value and system illustrates i phase. In Fig. 2(c), shell magnetization oscillates around the zero value and is delayed with respect to the external magnetic field and core magnetization does not follow the external magnetic field anymore, but instead of oscillating around a zero value, it oscillates around 0.5 value and system illustrates nm phase. These solutions do not depend on the initial values, seen in Fig. 2(a)-(c) explicitly.

*3.2. Thermal behaviors of the dynamic magnetizations*

The dynamic order parameters or the dynamic shell and core magnetizations as the time-averaged magnetization over a period of the oscillating magnetic field have given Eq. (5). With the combination of the Adams-Moulton predictor corrector and Romberg integration numerical methods, we solve Eq. (5) and examine the thermal behavior of dynamic magnetizations $M_\alpha$ ($\alpha$ = S (shell), C (core) and T (total)) for different values of system parameters. The thermal behaviors of dynamic magnetizations gives the dynamic phase transition (DPT) point and the type of the dynamic phase transition. Figs. 3(a)-(d) are presented for obtained numerical results of Eqs. (5), (6) and (11). In Fig. 3, $T_C$ and $T_t$ display the critical or the second-order phase transitions and the first-order phase transition temperatures, respectively. The $A_\alpha$, is dynamic hysteresis loop area and $C_\alpha$ is dynamic correlations. Fig. 3(a) shows the thermal behavior of dynamic magnetizations, dynamic hysteresis loop area and dynamic correlation for r = 1.0, $\Delta_S$ = 0.5, D = 0.0 and $h_0$ = 2.0 values. At zero temperature, $M_S$ = 1.0 and $M_C$ = 0.5 and with the increase of temperature they decrease to zero continuously; thus the system undergoes a second order phase transition from the ferrimagnetic (i) phase to the paramagnetic (p) phase at $T_C$ = 3.9. We have checked the stability of dynamical phase transition between the phases of the system by examine the dynamic hysteresis loop areas $A_\alpha$ and the dynamic correlations $C_\alpha$. The dynamic hysteresis loop areas and the dynamic correlations become a maximum and a minimum (negative) at the second-order phase transition temperature $T_C$, respectively. For r = -0.1, $\Delta_S$ = -0.5, D = 0.0 and $h_0$ = 1.0 values, the dynamic behavior of $M_\alpha$, $A_\alpha$ and $C_\alpha$ is obtained in Fig. 3(b). In this figure, $M_S$ and $M_C$ take 1.0 and -0.5 values at zero temperature, and they exhibits a continuous move to zero from these values. Hence, the system undergoes a second-order phase transition from the i phase to the p phase at $T_C$ = 1.08. The $A_S$ and $A_T$ dynamic hysteresis loop areas and the $C_S$ and $C_T$ dynamic correlations become a maximum and a minimum (negative) at the second-order phase transition temperature $T_C$, respectively. Moreover, the $A_C$ does not become a maximum and the $C_C$, become a maximum (positive) at $T_C$. Fig. 3(c) is plotted for r = 1.0, $\Delta_S$ = 0.0, D = 0.0 and $h_0$ = 3.7 values. In this figure, at zero temperature $M_S$ = 1.0 and $M_C$ = 0.5 and they decrease zero discontinuously as the temperature increases; hence, the system undergoes a first-order phase transition from the i phase to the p phase at $T_t$ = 0.53. Therefore, $T_t$ is the first-order phase transition temperature where the discontinuity or jump occurs. We also checked this dynamic discontinuous transition to investigate the thermal behavior of the dynamic hysteresis loop areas $A_\alpha$ and dynamic correlations $C_\alpha$, as seen in figure. As temperature increase from zero, the $A_\alpha$ and $C_\alpha$ increase from zero to a certain positive nonzero values, and $A_\alpha$ and $C_\alpha$ suddenly jump to the higher positive and lower negative values, respectively. Fig. 3(d) is obtained for r = 1.0, $\Delta_S$ = 0.0, D = 0.0 and $h_0$ = 3.4 and different initial values. We can see that the system undergoes two successive phase transitions; the first is a first-order phase transition from the p phase to the i phase at $T_t=0.78$, the second is a second-order one from the i phase to p phase at $T_C=1.36$. While the $A_\alpha$ decrease from zero as the temperature increase, the $C_\alpha$ increase from zero. They suddenly jump at $T_t=0.78$ values. Then, with the temperature increase the $A_\alpha$ become a maximum and $C_\alpha$ become a minimum (negative) at the second-order phase transition temperature $T_C = 1.36$.

## 3.3. Dynamic phase diagrams

Now, we can obtain the dynamic phase diagrams of the system. The dynamic phase diagrams are represented in the (h, T), (D, T), ($\Delta_S$, T) and (r, T) planes for different values of the physical parameters of the system. In Fig. 4, the solid and dashed lines stand for the second- and first-order phase transition lines, respectively. The dashed-dotted line illustrates the behavior of compensation temperatures. The dynamic tricritical point (TCP) is represented by a filled circle. The phase diagram in the (h, T) plane are illustrated in Fig. 4(a) for r = -1.0, $\Delta_S$ = -0.9 and D = -1.5 values. In Fig. 4(a), the system displays one dynamic TCP where signals the change from a first- to a second-order phase transition. Phase diagram contains i and p phases. Fig. 4(b) show the phase diagram in the (D, T) plane for the $h_0$ = 0.1, r = -1.0 and $\Delta_S$ = -0.9 values. As clearly seen from Fig. 4(b), the phase diagram include only second-order phase transition. The phase diagram contains i, p and nonmagnetic (nm) phases as well as compensation temperatures. One can clearly see that in low temperature and crystal field values, the system show nm phase. For $h_0$ = 0.1, r = -1.0 and D = -1.5 values, the phase diagram is plotted in the ($\Delta_S$, T) plane as seen in Fig. 4(c). Similar to Fig. 4(b), Fig. 4(c) also contains only second-order phase transition and compensation temperature. Phase diagram displays i and p phases. With the increase of surface exchange interaction parameter ($\Delta_S$), the phase transition temperature is increase. Finally, Fig. 4(d) is obtained to show the phase diagram in (r, T) plane for $h_0$ = 0.1, $\Delta_S$ = -0.9 and D = -1.5 values. The phase diagram contains i, p and nm phases, second-order phase transition lines as well as compensation temperatures.

## 3.4. The total magnetization behavior of mixed spin (1/2-1) HIN system

Fig. 5(a) displays the effect of the core-shell interfacial coupling on the total magnetization $M_T$ behavior. Fig. 5(a) is obtained for $h_0$ = 0.5, D = 0.0, $\Delta_S$ = -0.5 fixed values and r = -0.01, -0.5 and -1.0. In this figure, the P- and Q-type of compensation behaviors are obtained for r = -0.01, and -0.5 and -1.0 values, respectively. For the same values, the total dynamic hysteresis loop area $A_T$ and total dynamic correlations $C_T$ are obtained, as seen in Fig. 5(b) and 5(c), respectively. Fig. 6(a) is plotted for $h_0$ = 0.5, r = -1.0, D = 0.0 fixed values and for $\Delta_S$ = -0.99, -0.5, and 0.0 values to investigate the effect of the surface shell coupling on the $M_T$ behavior. For $\Delta_S$ = -0.99, and -0.5 and 0.0 values, the S- and Q-type of compensation behaviors are observed, respectively. In Fig. 6(b) and 6(c), the total dynamic hysteresis loop area $A_T$ and total dynamic correlations $C_T$ are presented, respectively. It can be easily seen from Fig. 6(b) that phase transition temperature is growing with the increase of the $\Delta_S$ values. Fig. 7(a) illustrates the influence of the crystal field on the $M_T$ behavior. Fig. 7(a) is obtained for $h_0$ = 0.5, r = -1.0, $\Delta_S$ = -0.5 fixed values and D = -1.0, -0.5 and 0.0. While the R-type is obtained for D = -1.0, the Q-type of compensation behaviors is observed for D = -0.5 and 0.0 values.

*3.5. The Compensation types of mixed spin (1/2-1) HIN system*

As known, the existence of the compensation temperature in a magnetic nanoparticle has important applications in the field of thermo-magnetic recording. In this purpose, we also studied the temperature variation of the total magnetization for various values of physical parameters of the system to obtain the compensation temperature and determine compensation types by using Eqs. (8) and (9). Fig. 8(a) shows the Q-type behaviors for the curve labeled $h_0 = 0.1$, $r = 0.75$, $\Delta_S = -0.5$ and $D = 0.25$. The R-type behavior is obtained in Fig. 8(b) for $h_0 = 0.1$, $r = 1.0$, $\Delta_S = -0.75$ and $D = -0.25$. Fig. 8(c) indicates the P-type behaviors for $h_0 = 0.1$, $r = 1.0$, $\Delta_S = -0.99$ and $D = 0.0$ values. For $h_0 = 0.5$, $r = -0.01$, $\Delta_S = -0.5$ and $D = 0.0$ values, the S-type behaviors is obtained as seen in Fig. 8(d). For $h_0 = 0.1$, $r = -0.75$, $\Delta_S = -0.99$ and $D = -1.0$ values, the N-type behaviors have observed as seen in Fig. 8(e). The Q-, R-, P- and N- types of compensations behaviors classified in the Néel theory [61] and S-type was obtained by Strecka [62]. It is also worth noting that recently the Q-, R-, S- and N-type [36] and the Q-, R-, N-, M-, P-, and S- type [35, 63] behaviors have been obtained in the mixed Ising nanoparticles and mixed hexagonal Ising nanowire systems, respectively.

## 4. Summary and Conclusion

Within the DEFT with correlations the dynamic phase transition points (DPTs), dynamic phase diagrams and dynamic compensation behaviors of the mixed spin (1/2-1) HIN system under a time oscillating longitudinal magnetic field were investigated. By utilizing the Glauber-type stochastic process, the EFT equations of motion for the average shell and core magnetizations are obtained for the system. We were presented the dynamic phase diagrams in the (h, T), (D, T), ($\Delta_S$, T) and (r, T) planes. Our results show that the dynamic phase diagrams contain the i, p and nm fundamental phases as well as TCP point and compensation temperature. The Q-, R-, P-, S- and N-types of compensation behaviors [35, 61, 62] have obtained in the system. We found that the dynamic behavior of the system strongly depends on the values of the interaction parameters. Finaly, although the equilibrium phase transitions of the nanosystems have been conspicuously studied, the dynamic or nonequilibrium properties of these systems have not been investigated considerably. Hence, we hope that our work contributes to close this shortcoming in the literature.

## References

[1] C. Alexiou, A. Schmidt, R. Klein, P. Hullin, C. Bergemann, W. Arnold, J. Magn. Magn. Mater., 252 (2002) 363.
[2] N. Sounderya, Y. Zhang, Recent Patents on Biomedical Engineering 1 (2008) 34.
[3] 22. G.V. Kurlyandskaya, M.L. Sanchez, B. Hernando, V.M. Prida, P. Gorria, M. Tejedor, Appl. Phys. Lett., 82 (2003) 3053.
[4] S. Nie, S.R. Emory, Science, 275 (1997) 1102.
[5] H. Zeng, J. Li, J.P. Liu, Z.L. Wang, S. Sun, Nature, 420 (2002) 395.
[6] D.W. Elliott, W.-X. Zhang, Environ. Sci. Tech., 35 (2001) 4922.
[7] J.E. Wegrowe, D. Kelly, Y. Jaccard, Ph. Guittienne, J.Ph. Ansermet, EPL, 45 (1999) 626.
[8] A. Fert, L. Piraux, J. Magn. Magn. Mater., 200 (1999) 338.
[9] R.H. Kodama, J. Magn. Magn. Mater., 200 (1999) 359.
[10] H.L. Su, G.B. Ji, S.L. Tang, Z. Li, B.X. Gu, Y.W. Du, Nanotechnology 16 (2005) 429; X. Lin, G. Ji, T. Gao, X. Chang, Y. Liu, H. Zhang Y. Du, Solid State Commun. 151

(2011) 1708.
[11] W. Chen, Z. Li, G.B. Ji, S.L. Tang, M. Lu, Y.W. Du, Solid State Commun. 133 (2005) 235;
J.Y. Chen, H.R. Liu, N. Ahmad, Y.L. Li, Z.Y. Chen, W.P. Zhou, X.F. Han, J. Appl. Phys. 109 (7) (2011) 07E157.
[12] H. Pan, B.H. Liu, J.B. Yi, C. Poh, S. Lim, J. Ding, Y.P. Feng, C. Huan, J.Y. Lin, J. Phys. Chem. B 109 (2005) 3094.
[13] K-H. Seong, J-Y. kim, J-J. Kim, S-C. Lee, S-R. Kim, U. Kim, T-E. Park, H-J. Choi, Nano Letters 7 (2007) 3366.
[14] B. Hamrakulov, I.S. Kim, M.G. Lee, et al., Trans. Nonferrous Metals Soc. China 19 (2009) 83.
[15] L. Zhang, Y. Zhang, J. Magn. Magn. Mater. 321 (2009) L15.
[16] N. Ahmad, J.Y. Chen, J. Iqbal, W.X. Wang, W.P. Zhou, X.F. Han, J. Appl. Phys. 109 (2011) 07A331;
S. Pal, S. Saha, D. Polley, A. Barman, Solid State Commun., in press.
[17] J.P. Xu, Z.Z. Zhang, B. Ma, Q.Y. Jin, J. Appl. Phys. 109 (2011) 07B704.
[18] C. Rousse, P. Fricoteaux, J. Mater. Sci. 46 (2011) 6046.
[19] Z. H. Yang, Z. W. Li, L. Liu, L. B. Kong, J. Magn. Magn. Mater. 323 (2011) 2674.
[20] T. Kaneyoshi, J. Magn. Magn. Mater. 322 (2010) 3410; 322 (2010) 3014; 323 (2011) 3014; Physica A 390 (2011) 3697; Phys. Stat. Sol. B 248 (2011) 250.
[21] M. Keskin, N. Şarlı, B. Deviren, Solid State Commun. 151 (2011) 2483.
[22] M. Boughrara, M. Kerouad, A. Zaim, J. Magn. Magn. Mater. 360 (2014) 222; 368 (2014) 169.
[23] Y. Kocakaplan, E. Kantar, Chinese Phys. B 23 (2014) 046801.
[24] E. Kantar, M. Keskin, J. Magn. Magn. Mater. 349 (2014) 165.
[25] A. Feraoun, A. Zaim, M. Kerouad, Physica B 445 (2014) 74.
[26] T. Kaneyoshi, E.F. Sarmento, I.F. Fittipaldi, Phys. Stat. Sol. B 150 (1988) 261.
[27] T. Kaneyoshi, J.C. Chen, J. Magn. Magn. Mater. 98 (1991) 201.
[28] J.A. Plascak, Physica A 198 (1993) 665.
[29] M. Keskin, M. Ertaş, J. Stat. Phys. 139 (2010) 333.
[30] G.M. Zhang, C.Z. Yang, Phys. Rev. B 48 (1993) 9452.
[31] G.M. Buendia, M.A. Novotny, J. Zhang, in: D.P. Landau, K.K. Mon, H.B. Schuttler (Eds.), Springer Proceeds in Physics 78, Computer Simulations in Condensed Matter Physics, Vol. VII, Springer, Heidelberg, 1994, p. 223.
[32] S.L. Yan, L. Liu, J. Magn. Magn. Mater. 312 (2007) 285.
[33] Y.F. Zhang, S.L. Yan, Solid State Commun. 146 (2008) 478; Phys. Lett. A 372 (2008) 2696.
[34] W. Jiang, G.Z. Wei, A. Du, J. Magn. Magn. Mater. 250 (2002) 49.
[35] C. Ekiz, J. Strecka, M. Jascur, Cent. Eur. J. Phys. 7 (2009) 509.
[36] E. Kantar, Y. Kocakaplan, Solid State Commun. 177 (2014) 1.
[37] Y. Kocakaplan, E. Kantar, Eur. Phys. J. B 87 (2014) 135.
[38] A. Zaim, M. Kerouad, Y. E. Amraoui, J. Magn. Magn. Mater. 321 (2009) 1077.
[39] N. Şarlı, Physica B 411 (2013) 12.
[40] O. Canko, A. Erdinç, F. Taşkın, A. F. Yıldırım, J. Magn. Magn. Mater. 324 (2012) 508.
[41] T. Tomé, M.J. de Oliveira, Phys. Rev. A 41 (1990) 4251.
[42] G.M. Buendía, E. Machado, Phys. Rev. E 58 (1998) 1260.
[43] B.K. Chakrabarti, M. Acharyya, Rev. Mod. Phys. 71 (1999) 847.
[44] S.W. Sides, P.A. Rikvold, M.A. Novotny, Phys. Rev. E 59 (1999) 2710.
[45] M. Keskin, O. Canko, U. Temizer, Phys. Rev. E 72 (2005) 036125.

[46] M. Keskin, E. Kantar, O. Canko, Phys. Rev. E 77 (2008) 051130.
[47] M. Keskin, E. Kantar, J. Magn. Magn. Mater. 322 (2010) 2789.
[48] M. Ertaş, B. Deviren, M. Keskin, Phys. Rev. E 86 (2012) 051110.
[49] M. Ertaş, Y. Kocakaplan, M. Keskin, J. Magn. Magn. Mater. 348 (2013) 113.
[50] Q. Jiang, H.-N. Yang, G.C. Wang, Phys. Rev. B 52 (1995) 14911.
[51] J.S. Suen, J.L. Erskine, Phys. Rev. Lett. 78 (1997) 3567-3570, May. 1997.
[52] D.T. Robb, Y.H. Xu, O. Hellwig, J. McCord, A. Berger, M.A. Novotny, P.A. Rikvold, Phys. Rev. B 78 (2008) 134422.
[53] N. Gedik, D.-S. Yang, G. Logvenov, I. Bozovic, and A.H. Zewail, Science 316 (2007) 425.
[54] K. Kanuga, M. Cakmak, Polymer 48 (2007) 7176.
[55] B. Deviren, M. Ertaş, M. Keskin, Phys. Scripta 85 (2012) 055001.
[56] B. Deviren, E. Kantar, M. Keskin, J. Magn. Magn. Mater. 324 (2012) 2163.
[57] M Ertaş, Y Kocakaplan, Phys. Lett. A 378 (2014) 845.
[58] E Kantar, M Ertaş, M Keskin, J. Magn. Magn. Mater.361 (2014) 61
[59] V.S. Leite, W. Figueirredo, Physica A 350 (2005) 379.
[60] Y. Yüksel, E. Vatansever, H. Polat, J. Phys.: Condens. Matter 24 (2012) 436004.
[61] L. Néel, Ann. Phys. 3 (1948) 137.
[62] J. Strečka, Physica A, 360 (2006) 379.
[63] Y. Kocakaplan, E. Kantar, and M. Keskin, Eur. Phys. J. B 86 (2013) 420.

**List of Figure Captions**

**Fig. 1.** (Color online) Schematic presentation of hexagonal Ising nanowire. The blue and red spheres indicate magnetic atoms at the surface shell and core, respectively.

**Fig. 2.** (Color online) Time variations of the core and shell magnetizations ($m_C$ and $m_S$):

(a) Paramagnetic phase (p), $r = 1.0$, $\Delta_S = 0.5$, $D = 0.0$ and $h_0 = 2.0$ and $T = 4.5$.

(b) Ferrimagnetic phase (f), $r = -0.1$, $\Delta_S = -0.5$, $D = 0.0$ and $h_0 = 1.0$, and $T = 0.5$.

(c) Nonmagnetic phase (nm), $r = -0.25$, $\Delta_S = -0.9$, $D = -1.5$ and $h_0 = 0.1$, and $T = 0.1$.

**Fig. 3.** (Color online) Thermal behaviors of the dynamic core and shell magnetizations with the various values of r and $\Delta_S$.

(a) $r = 1.0$, $\Delta_S = 0.5$, $D = 0.0$ and $h_0 = 2.0$

(b) $r = -0.1$, $\Delta_S = -0.5$, $D = 0.0$ and $h_0 = 1.0$

(c) $r = 1.0$, $\Delta_S = 0.0$, $D = 0.0$ and $h_0 = 3.7$

(d) $r = 1.0$, $\Delta_S = 0.0$, $D = 0.0$ and $h_0 = 3.4$

**Fig. 4.** The dynamic phase diagrams in the (h, T), (D, T), ($\Delta_S$, T) and (r, T) planes of the hexagonal Ising nanowire. The solid and dashed lines stand for the second- and first-order phase transition lines, respectively. The dashed-dotted line illustrates the behavior of compensation temperatures. The dynamic tricritical point (TCP) is represented by a filled circle.

(a) $r = -1.0$, $\Delta_S = -0.9$ and $D = -1.5$

(b) $r = -1.0$, $\Delta_S = -0.9$ and $h_0 = 0.1$

  (c) $r = -1.0$, $h_0 = 0.1$ and $D = -1.5$

  (d) $\Delta_S = -0.9$, $h_0 = 0.1$ and $D = -1.5$

**Fig. 5.** (Color online) For $h_0 = 0.5$, $D = 0.0$, $\Delta_S = -0.5$ fixed values and $r = -0.01$, $-0.5$ and $-1.0$ values

  (a) Total dynamic magnetization

  (b) Total dynamic correlations

  (c) Total dynamic hysteresis loop area.

**Fig. 6.** Same as with Fig. 5, but for $h_0 = 0.5$, $r = -1.0$, $D = 0.0$ fixed values and for $\Delta_S = -0.99$, $-0.5$, and $0.0$ values

  (a) Total dynamic magnetization

  (b) Total dynamic correlations

  (c) Total dynamic hysteresis loop area.

**Fig. 7.** Same as with Fig. 5, but for $h_0 = 0.5$, $r = -1.0$, $\Delta_S = -0.5$ fixed values and $D = -1.0$, $-0.5$ and $0.0$. values

  (a) Total dynamic magnetization

  (b) Total dynamic correlations

  (c) Total dynamic hysteresis loop area.

**Fig. 8.** The type of compensation behaviors for:

  (a) $h_0 = 0.1$, $r = 0.75$, $\Delta_S = -0.5$ and $D = 0.25$.

  (b) $h_0 = 0.1$, $r = 1.0$, $\Delta_S = -0.75$ and $D = -0.25$.

  (c) $h_0 = 0.1$, $r = 1.0$, $\Delta_S = -0.99$ and $D = 0.0$.

  (d) $h_0 = 0.5$, $r = -0.01$, $\Delta_S = -0.5$ and $D = 0.0$.

  (e) $h_0 = 0.1$, $r = -0.75$, $\Delta_S = -0.99$ and $D = -1.0$.

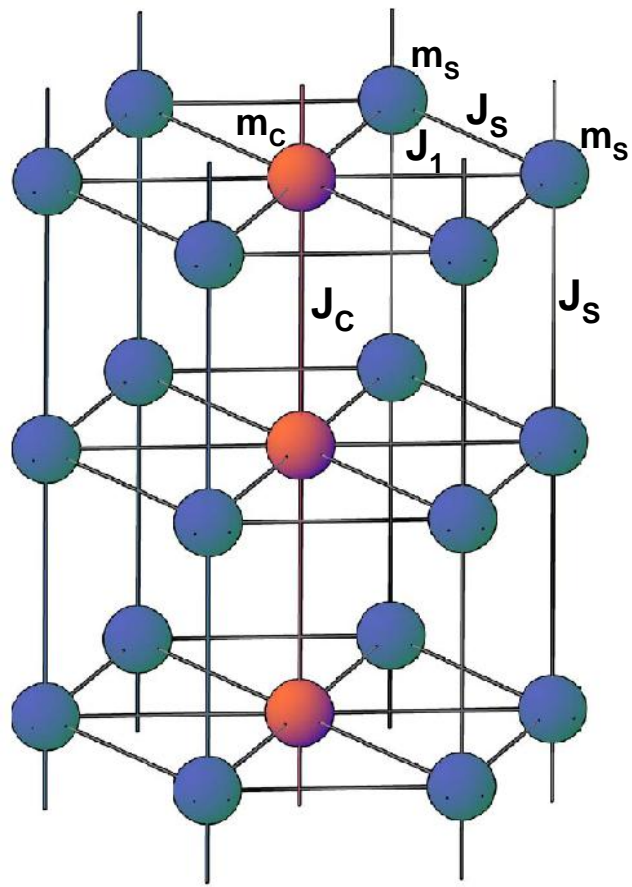

Fig. 1

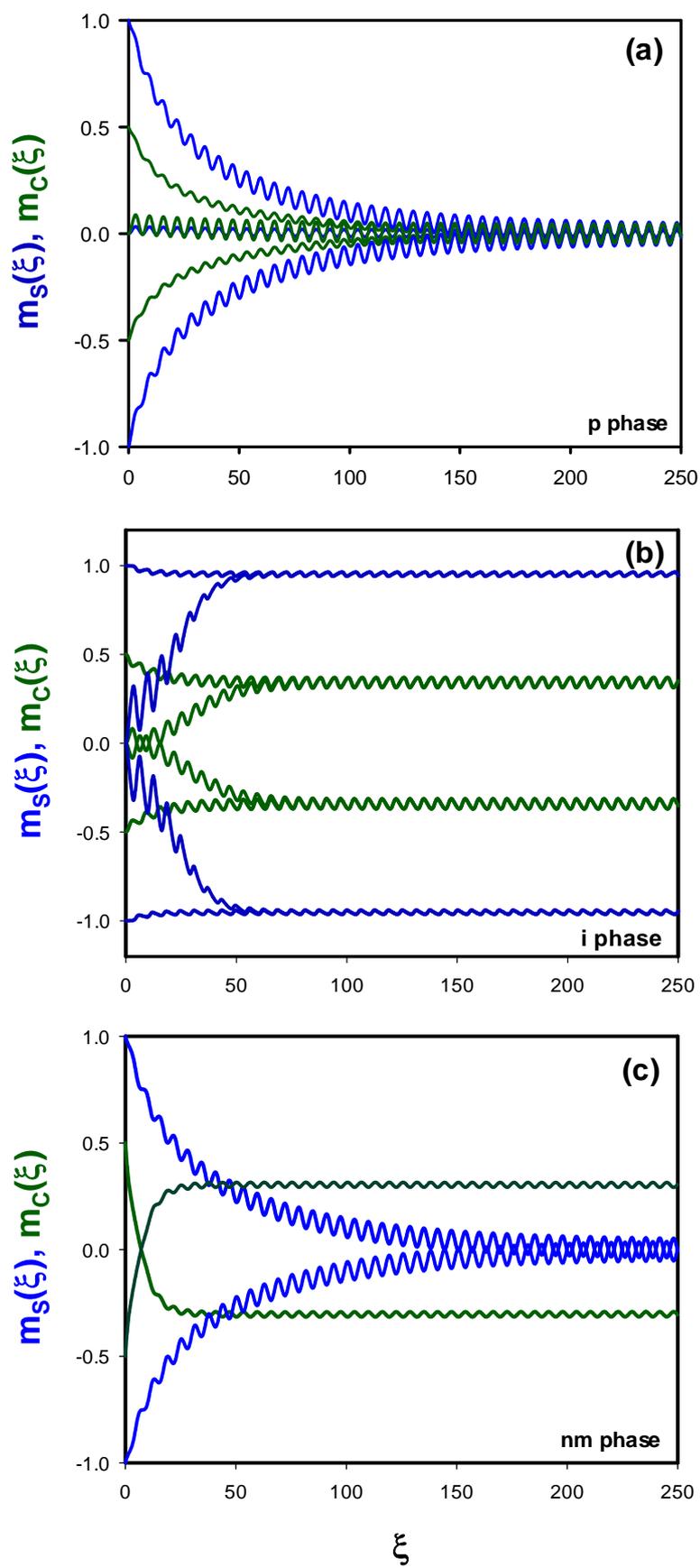

Fig. 2

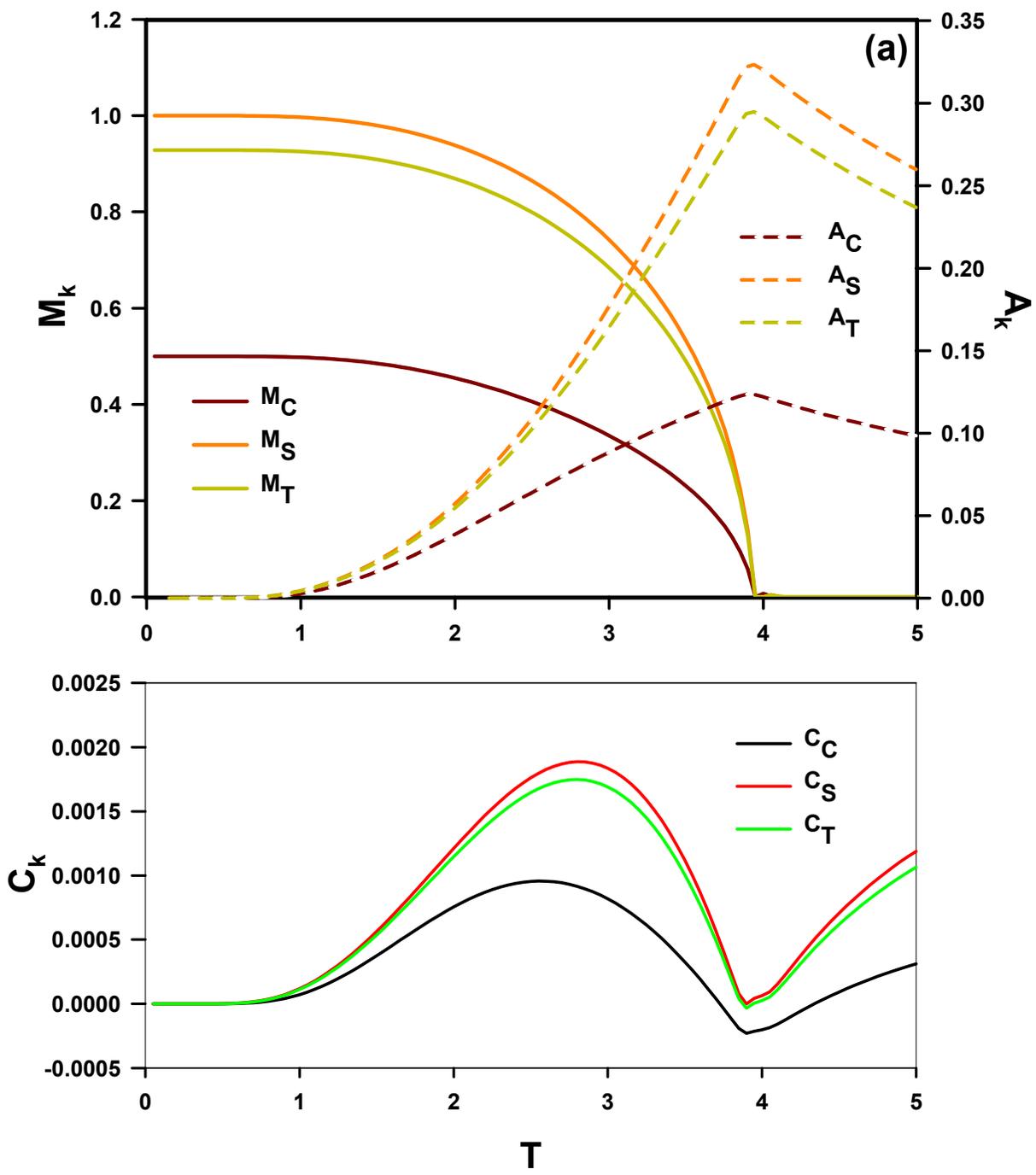

Fig. 3a

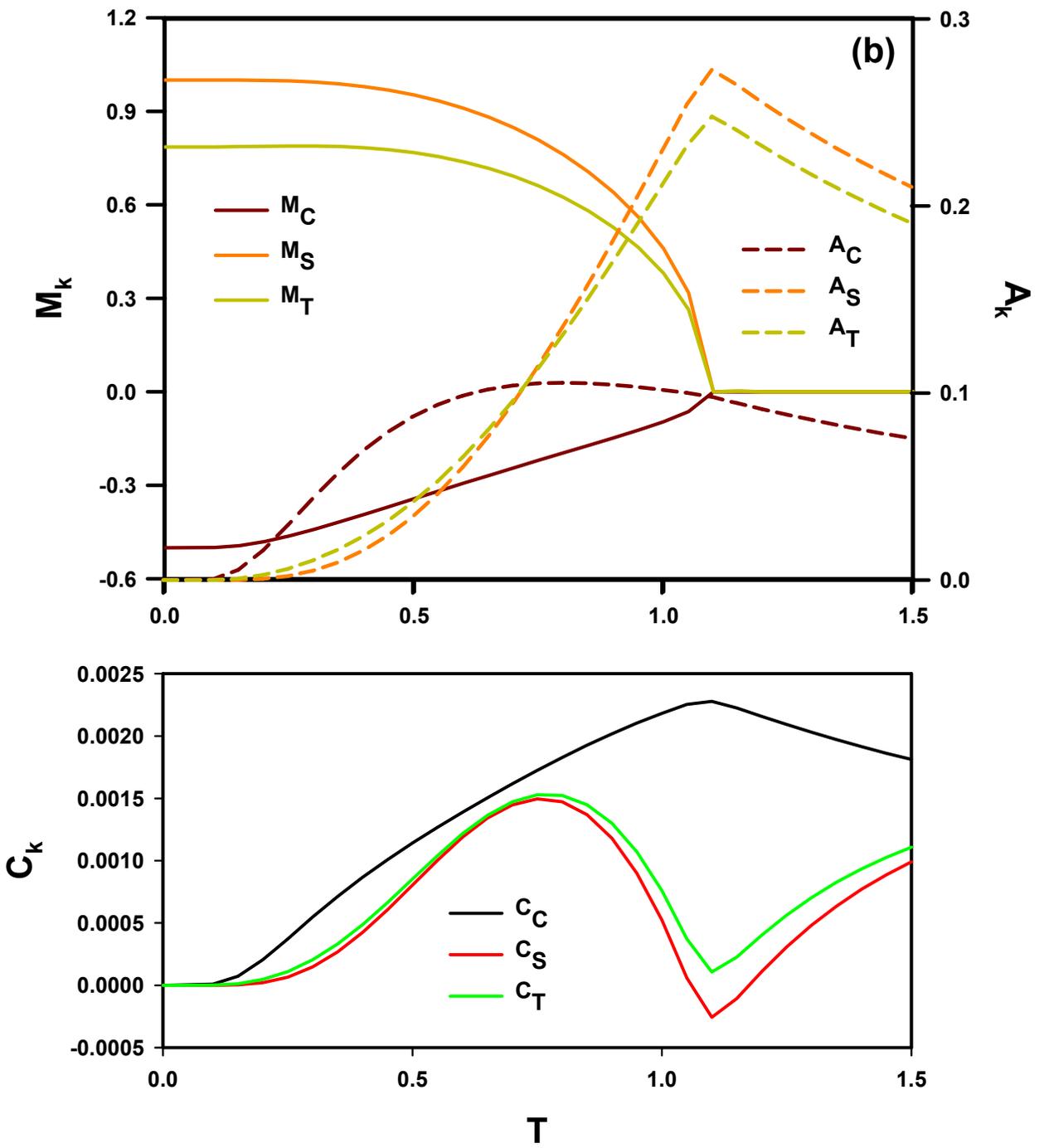

Fig. 3b

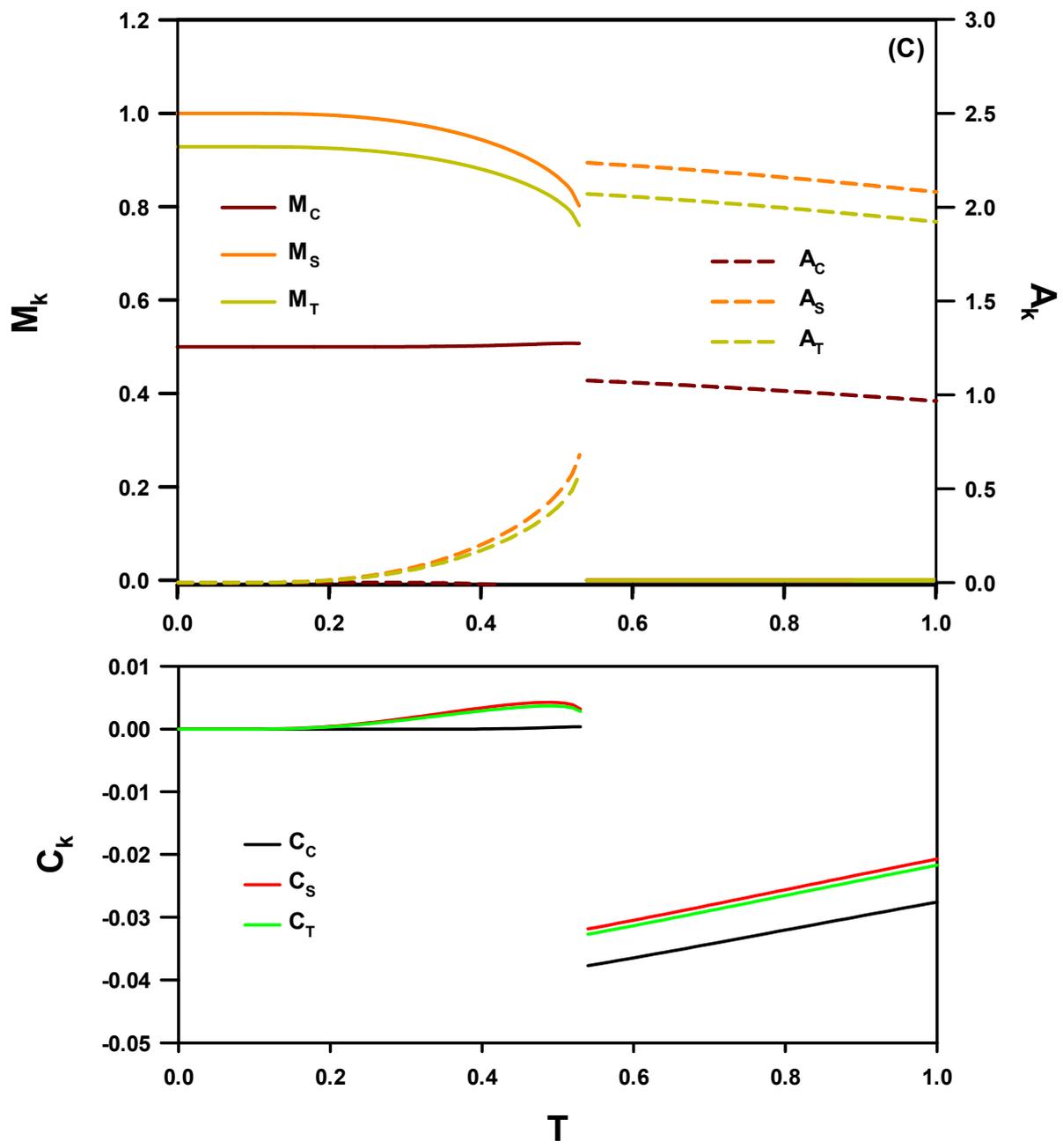

Fig. 3c

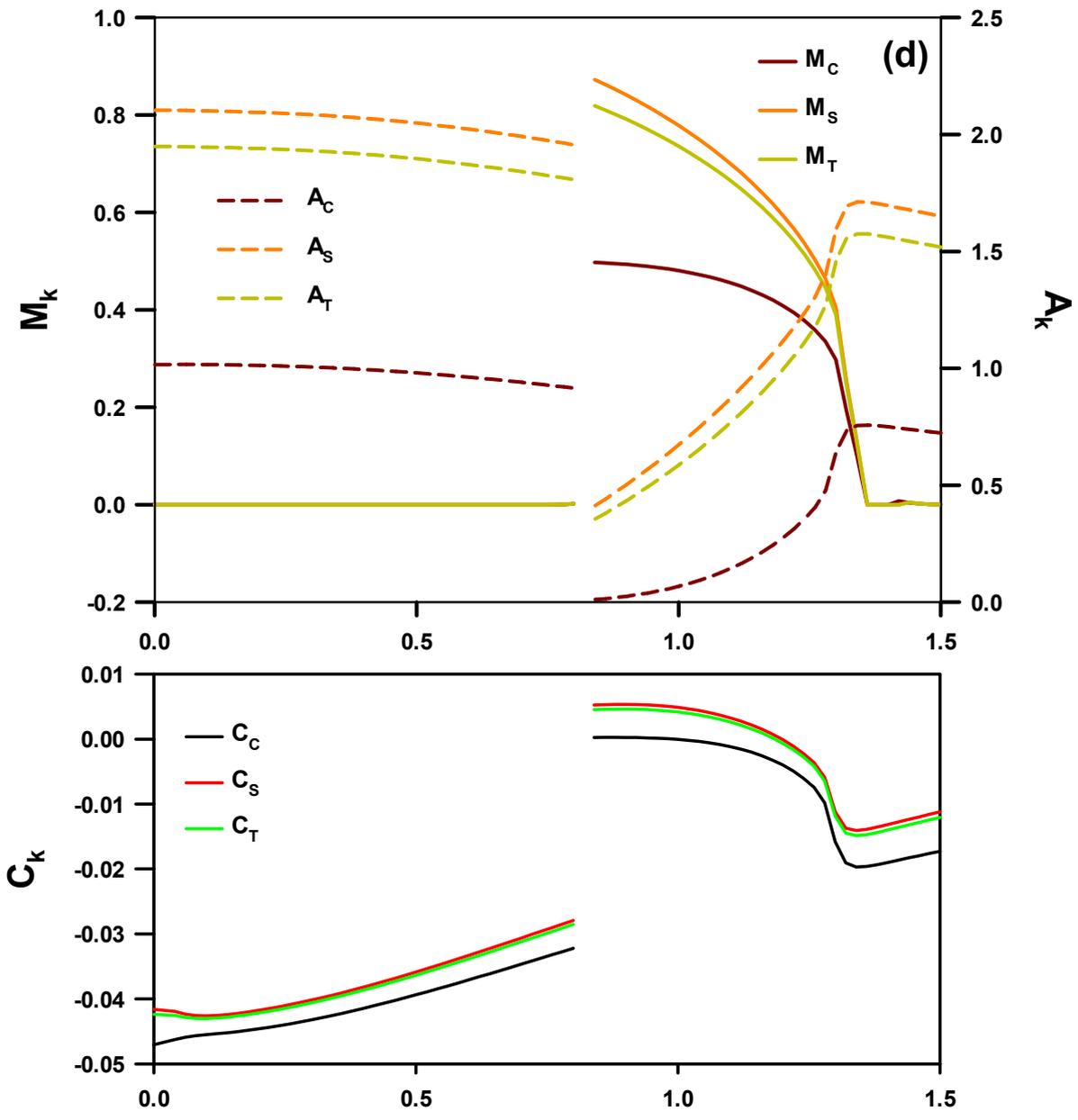

Fig. 3d

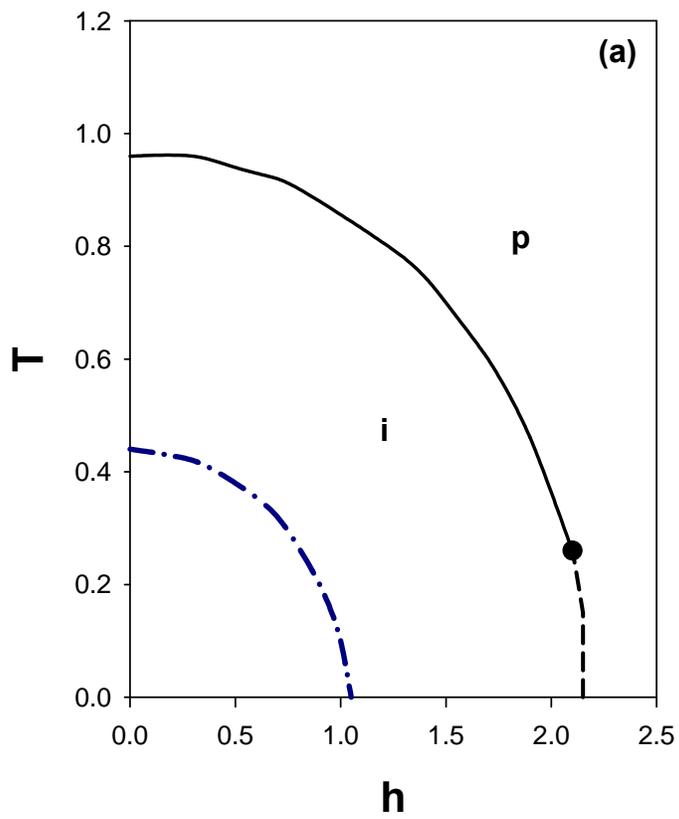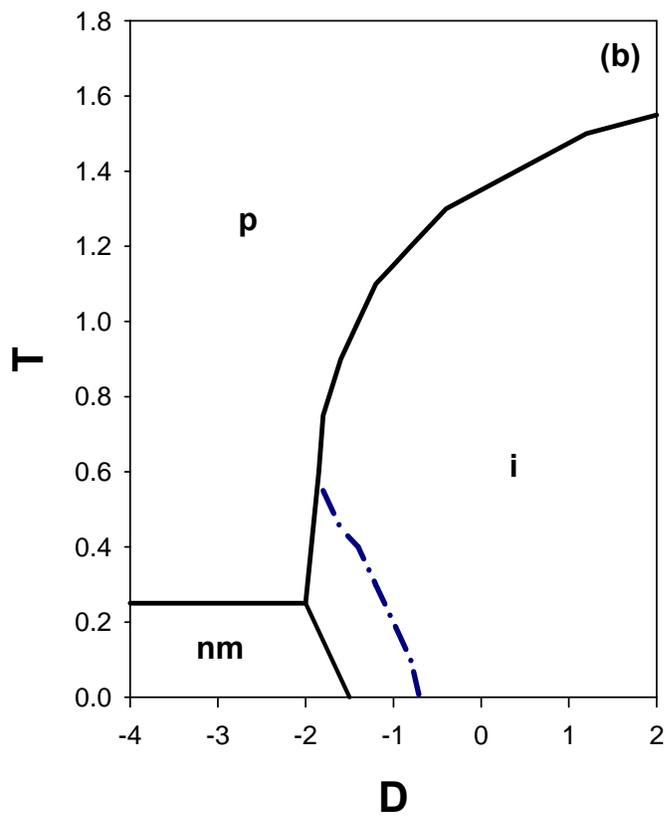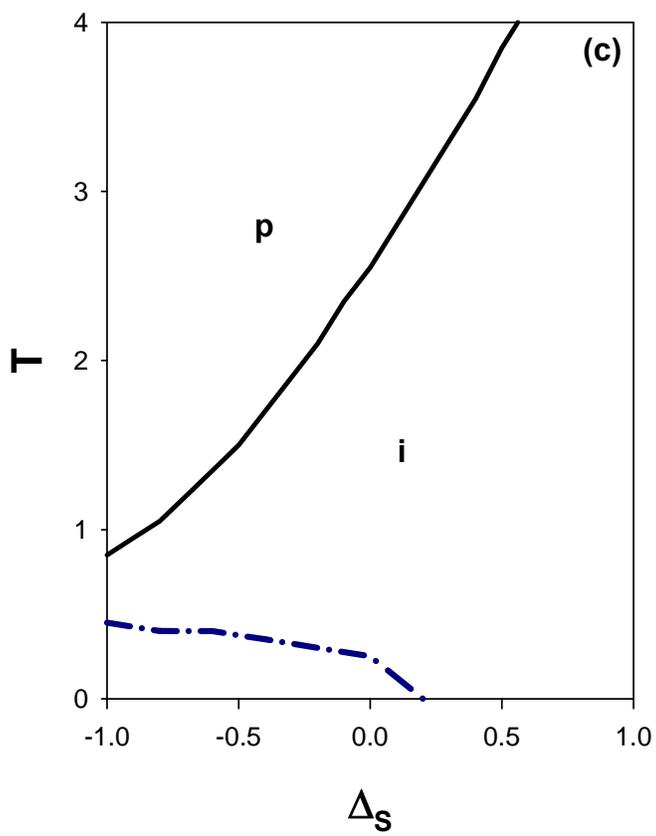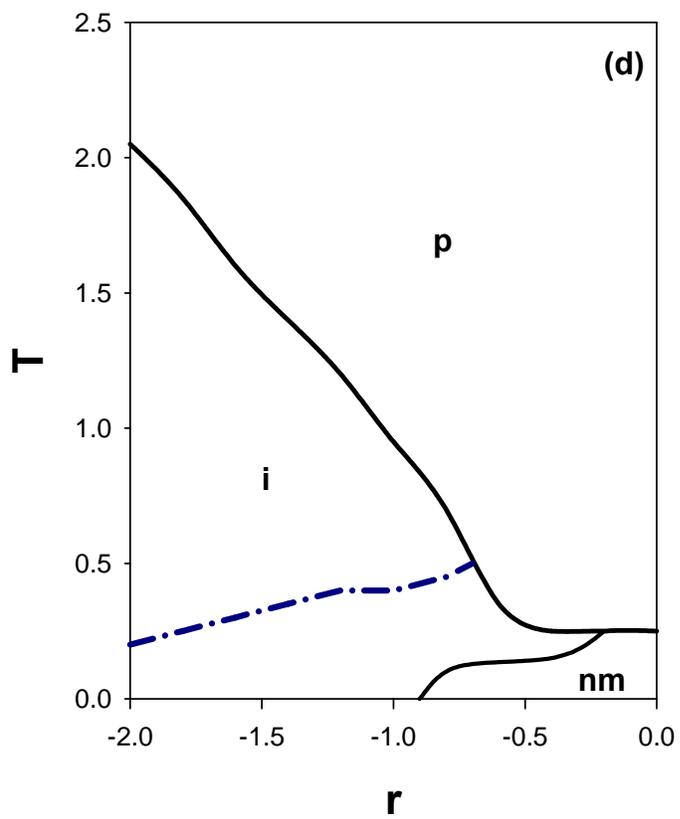

Fig. 4

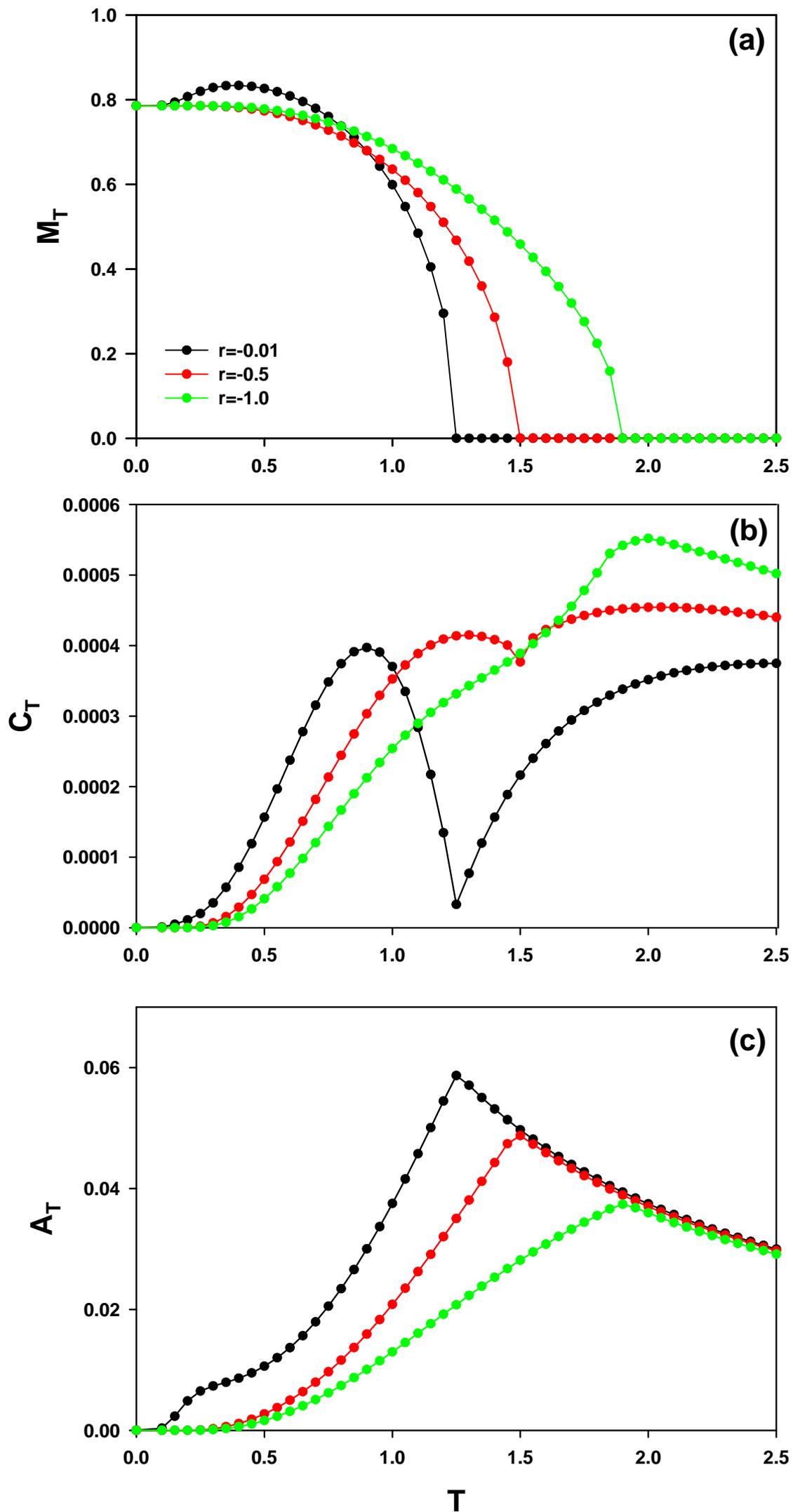

Fig. 5

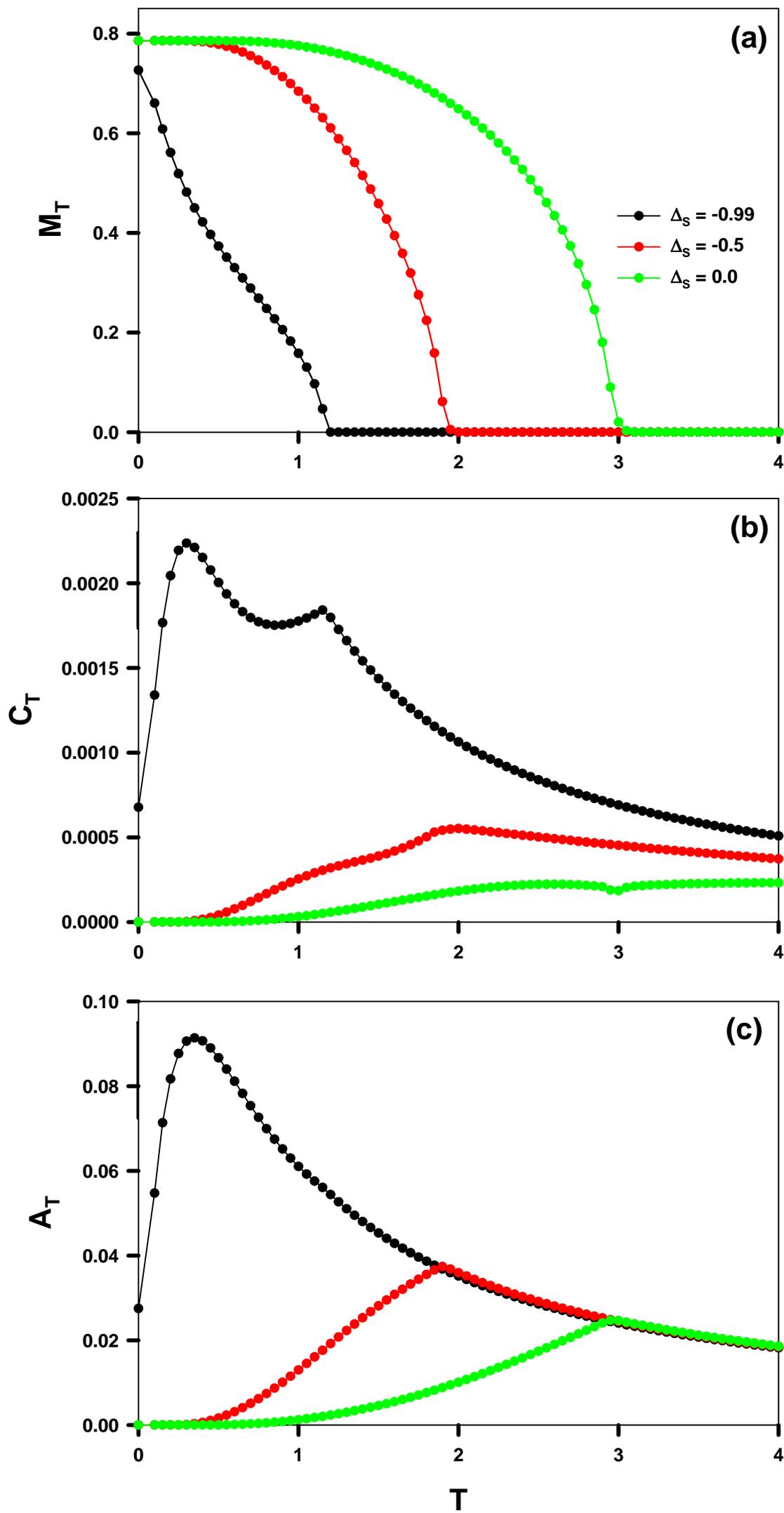

Fig. 6

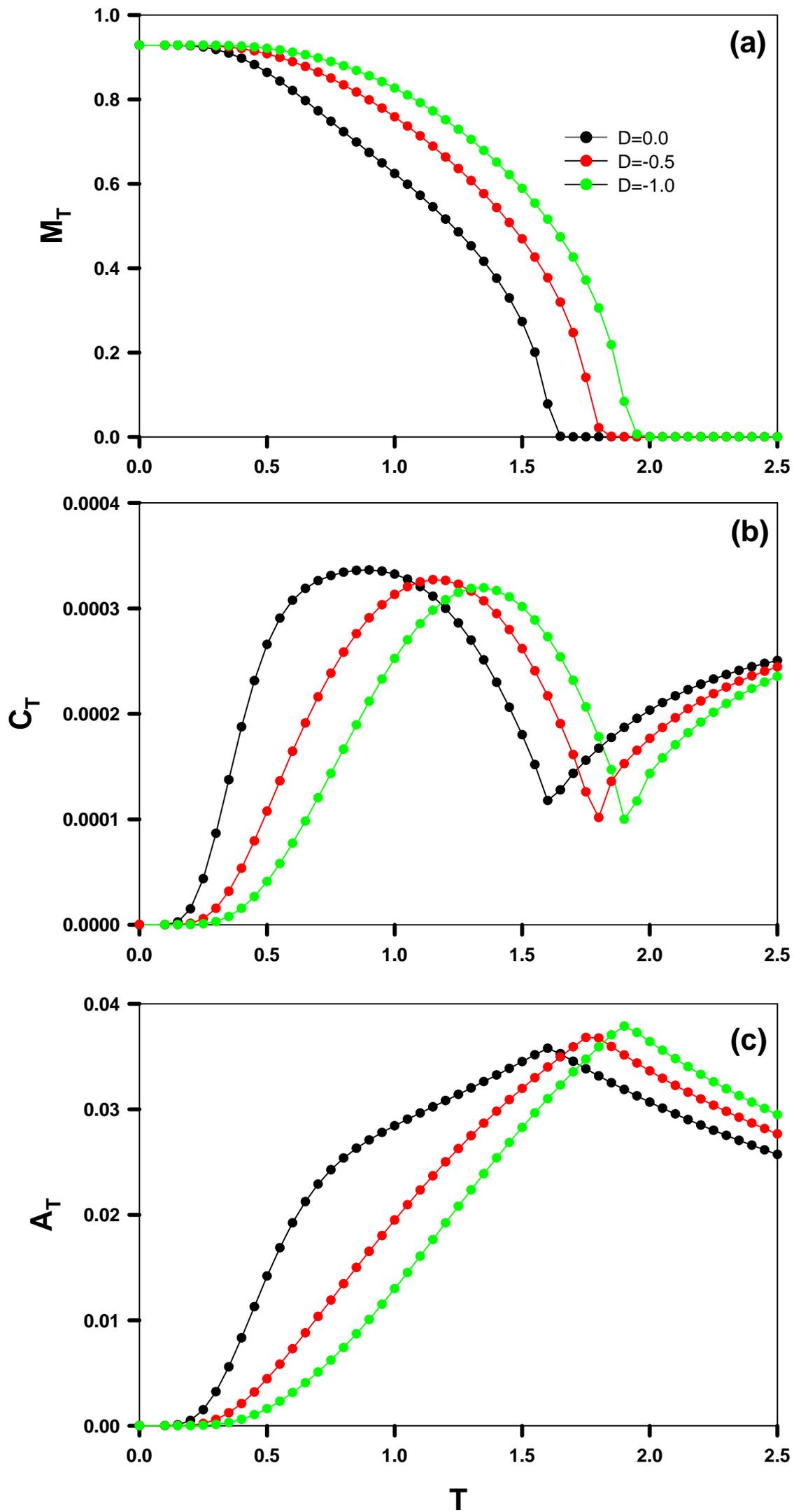

Fig. 7

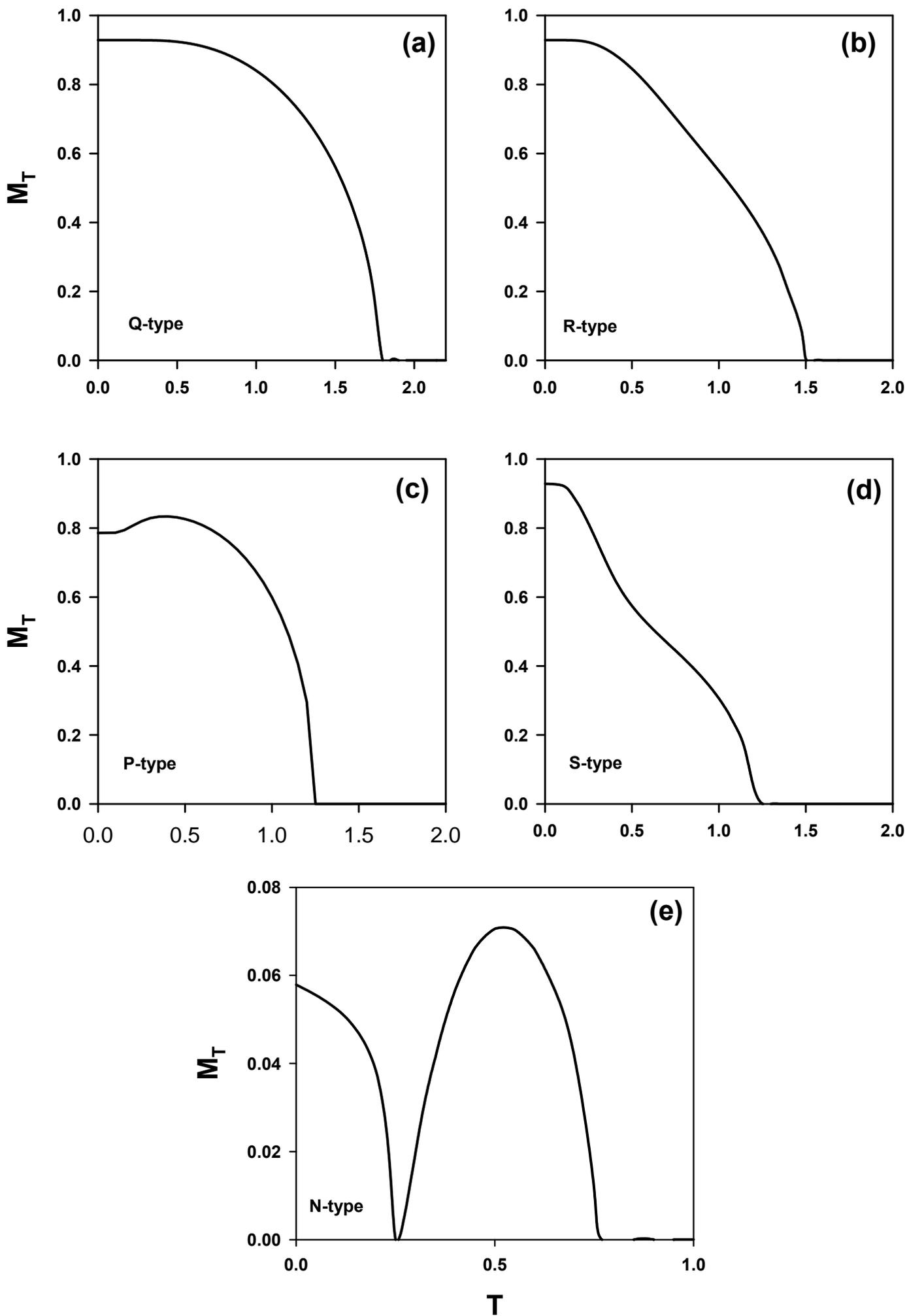

Fig. 8